\documentclass{article}

\usepackage{arxiv}

\usepackage[utf8]{inputenc} 
\usepackage[T1]{fontenc}    
\usepackage{hyperref}       
\usepackage{url}            
\usepackage{booktabs}       
\usepackage{amsfonts}       
\usepackage{nicefrac}       
\usepackage{microtype}      
\usepackage{lipsum}		
\usepackage{graphicx}
\usepackage{natbib}
\usepackage{doi}

\usepackage{amsmath}
\usepackage{nicefrac}
\usepackage{xcolor}
\usepackage{bm}
\renewcommand{\vec}[1]{\bm{#1}}
\newcommand{\ten}[1]{\bm{#1}}
\usepackage{appendix}

\title{Dirichlet and Neumann boundary conditions in a Lattice Boltzmann Method for Elastodynamics}

\date{August 8, 2022}	

\author{ \href{https://orcid.org/0000-0003-2189-945X}{\includegraphics[scale=0.06]{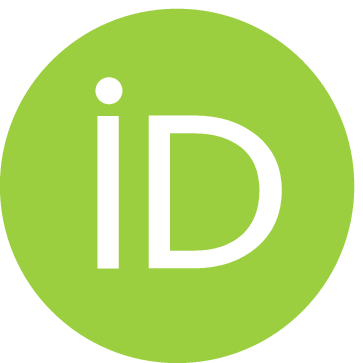}\hspace{1mm}Erik Faust}\thanks{Corresponding author.} \\
	Lehrstuhl für Technische Mechanik\\
	Technische Universität Kaiserslautern\\
	Postfach 3049, D-67653 Kaiserslautern \\
	\texttt{efaust@rhrk.uni-kl.de} \\
	\And
	\href{https://orcid.org/0000-0002-0839-6519}{\includegraphics[scale=0.06]{orcid.eps}\hspace{1mm}Alexander Schlüter} \\
	Lehrstuhl für Technische Mechanik\\
	Technische Universität Kaiserslautern\\
	Postfach 3049, D-67653 Kaiserslautern \\
	\texttt{aschluet@rhrk.uni-kl.de} \\
	\And
	\href{https://orcid.org/0000-0003-4819-2198}{\includegraphics[scale=0.06]{orcid.eps}\hspace{1mm}Henning Müller} \\
	Institut für Mechanik\\
	Technische Universität Darmstadt\\
	Franziska-Braun-Straße~7, D-64287, Darmstadt \\
	\texttt{henning.mueller@tu-darmstadt.de} \\
	\And
	{\hspace{1mm}Ralf Müller} \\
	Institut für Mechanik\\
	Technische Universität Darmstadt\\
	Franziska-Braun-Straße~7, D-64287, Darmstadt \\
	\texttt{ralf.mueller@mechanik.tu-darmstadt.de} \\
}

\renewcommand{\shorttitle}{Boundary conditions in a solid LBM}

\hypersetup{
pdftitle={Boundary conditions in a solid LBM},
pdfsubject={q-bio.NC, q-bio.QM},
pdfauthor={Erik Faust, Alexander Schlüter, Henning Müller, Ralf Müller},
pdfkeywords={Lattice Boltzmann Method, Elastodynamics, Boundary Conditions, Transient Solid Simulation},
}

\begin{document}
\maketitle

\begin{abstract}
    Recently, Murthy et al. \cite{murthy} and Escande et al. \cite{escande} adopted the Lattice Boltzmann Method (LBM) to model the linear elastodynamic behaviour of isotropic solids. 
    The LBM is attractive as an elastodynamic solver because it can be parallelised readily and lends itself to finely discretised dynamic continuum simulations, allowing transient phenomena such as wave propagation to be modelled efficiently. 
    
    This work proposes simple local boundary rules which approximate the behaviour of Dirichlet and Neumann boundary conditions with an LBM for elastic solids. Both lattice-conforming and non-lattice-conforming, curved boundary geometries are considered.
    
    For validation, we compare results produced by the LBM for the sudden loading of a stationary crack with an analytical solution from \cite{freund}. 
    Furthermore, we investigate the performance of the LBM for the transient tension loading of a plate with a circular hole, using Finite Element (FEM) simulations as a reference.
\end{abstract}

\keywords{Lattice Boltzmann Method \and Elastodynamics \and Boundary Conditions \and Transient Solid Simulation}

\section{Introduction}

As a viable simulation method, the Lattice Boltzmann Method (LBM) was first established in the context of fluid mechanics – in which the LBMs' distribution functions are most directly subject to physical interpretation, and technologically relevant transient phenomena are commonplace \cite{escande}. However, solid mechanics also features its share of such phenomena – stress wave superposition and dynamic overshoots in the displacement under dynamic loading can result in significantly higher stresses than those predicted by quasistatic calculations – with far-reaching consequences for mining and crash-proofing, for example \cite{hashemi}. Since analytical calculations are often impossible and efficiency is a bottleneck for numerical computation in these cases, researchers have recently begun developing LBM algorithms for solid simulation: Marconi et al. \cite{marconi} approximated crack propagation with one such algorithm, and O'Brien et al. \cite{obrien} based another on a wave equation for Poisson solids to model wave propagation. More recently, Murthy et al. \cite{murthy} and Escande et al. \cite{escande} extended the solid LBM to the more general class of isotropic, linear elastic solids. Given further developments, it may become feasible to perform coupled fluid-structure simulations using the LBM as both a fluid and a solid solver.

Thus, the groundwork for LBM-based transient solid simulations has been laid, but much work remains to be done if LBM algorithms are to become viable tools for applied solid mechanics, or for fluid-structure simulation. While efficient, these algorithms are not stable for all relevant combinations of material parameters and loads \cite{escande}. Additionally, only rudimentary boundary conditions – for periodic, free, and fixed boundaries, all of which must be lattice-conforming – may be modelled currently \cite{escande}. 

This work takes aim at the latter obstacle. Simple local bounce-back-type boundary rules which may be used to model arbitarily valued Dirichlet and Neumann boundaries in the solid case are presented for both lattice-conforming and arbitrary geometries. 

Before the formulation of these boundary rules are addressed, however, section~\ref{s:mech} discusses the reformulation of the underlying solid mechanical equations required by the LBM. The Lam\'e-Navier equation is restated as a moment chain and boundary conditions are rephrased accordingly. Section~\ref{s:lbm} outlines the Lattice Boltzmann Method for solids briefly, and section~\ref{s:bound} presents the novel boundary rules. In section~\ref{s:valid}, we validate the LBM algorithm and the boundary rules against one analytical and one numerical benchmark problem. 
While the modified boundary rules are easily generalised, for ease of visualisation and validation we limit the discussion to two-dimensional (plane strain) problems here.

\section{A moment chain formulation for linear elastodynamics}\label{s:mech}

For sufficiently small displacements and displacement derivatives\footnote{See e.g. \cite[p.3]{landau}. Quadratic (and higher) products of displacement derivatives are neglected.}, the Lam\'e-Navier equation
\begin{equation}\label{eq:navier}
    \rho \partial_t^2 u_\alpha = (\lambda + \mu) \partial_\alpha \partial_\beta u_\beta + \mu \partial_\beta \partial_\beta u_\alpha + F_\alpha
\end{equation}
approximates the temporal and spatial evolution of the displacement $\vec{u}$ in isotropic, elastic continua \cite[p.39]{becker}. Here, $\rho$ denotes the (current) mass density, $\lambda$ and $\mu$ are the Lam\'e parameters, and $\vec{F}$ is a bulk force. The del symbols indicate the partial derivatives with respect to time ($\partial_t \cdot = \nicefrac{\partial \cdot}{\partial t}$) and with respect to the spatial coordinates ($\partial_\alpha \cdot = \nicefrac{\partial \cdot}{\partial x_\alpha}$),
respectively. Furthermore, Einstein summation convention is used. Equation (\ref{eq:navier}) can be viewed as the small strain limit of the equations of motion in the current configuration, $\vec{x} \in \Omega$.

For a solution to be determined on a bounded domain $\Omega$, initial conditions as well as boundary conditions (BCs) for the displacement (Dirichlet boundary conditions)
\begin{equation}\label{eq:bcdir}
    u_\alpha(\vec{x},t) = u_\alpha^*(\vec{x},t), \: \vec{x} \in \partial \Omega_u\,,
\end{equation}
and/or for the surface traction (Neumann boundary conditions)
\begin{equation}\label{eq:bcneu}
    \sigma_{\alpha \beta}(\vec{x},t) n_\beta = t_\alpha^*(\vec{x},t), \: \vec{x} \in \partial \Omega_t\,,
\end{equation}
on regions $\partial \Omega_u$ and $\partial \Omega_t$, respectively, must further be defined, where ${\partial \Omega_u \cup \partial \Omega_t = \partial \Omega}$ \cite[p.7]{poruchikov}.

While boundary conditions are easily specified, accounting for them in a simulation method is often non-trivial, and the LBM makes no exception~\cite[p.155]{krueger}. Moreover, because boundary conditions are central to engineering problems -- much of the information relevant to an engineering design process is contained in boundary conditions -- their importance is hard to overstate.

\subsection{A moment chain of conservation laws}

Under certain smallness assumptions\footnote{We have not found a thorough discussion of these assumptions in the literature. An in-depth investigation of the solid-mechanical assumptions underlying the solid LBM would be interesting, but would also expand the scope of this publication considerably. Broadly speaking, terms beyond the leading order of smallness in the displacement derivatives are neglected in each equation in (\ref{eq:chainmod}).}, the Lam\'e-Navier equation (\ref{eq:navier}) can be rephrased as a moment chain \cite{murthy}
\begin{align}
    & \partial_t \rho + \partial_\alpha j_\alpha = 0\,,\nonumber \\
    & \partial_t j_\alpha + \partial_\beta P_{\alpha \beta} = F_\alpha + \frac{\mu-\lambda}{\rho} \partial_\alpha \rho\,, \nonumber \\
    & \partial_t P_{\alpha \beta} + \partial_\gamma \Big ( \frac{\mu}{\rho} (
    j_\alpha \delta_{\beta \gamma} + j_\beta \delta_{\alpha \gamma} + j_\gamma \delta_{\beta \alpha} )
    \Big ) = 0\,. \label{eq:chainmod}
\end{align}
Therein, $\vec{j}=\rho \partial_t \vec{u}$ denotes the linear momentum density and $\ten{P}$ is a stress tensor which is generally not equivalent to the established Cauchy stress $\ten{\sigma}$.
In (\ref{eq:chainmod}), the first equation accounts for the conservation of mass, the second for the balance of linear momentum, and the third for the material law.
Note that this moment chain is somewhat inconsistent with formal small strain theory, which assumes the current and reference configurations to coincide and the current density $\rho$ and the reference density $\rho_0$ to be identical \cite[p.4]{poruchikov}. 
However, the difference this introduces is of third order in the displacement derivatives and becomes negligible for infinitesimal deformations.

Mathematically, each equation in (\ref{eq:chainmod}) contains the time derivative of a tensor of order $N$ and the divergence of a tensor of order $N+1$ on the left, as well as an $N$-th order source term on the right hand side.  Farag et al. \cite{farag} demonstrated that the LBM acts as a second-order Crank-Nicholson scheme for moment chains with precisely this structure, with the tensors appearing on the left hand side of (\ref{eq:chainmod}) -- $\rho$, $j$, and $P$ -- being accounted for as moments in the LBM. 
This explains why Murthy et al. \cite{murthy} and Escande et al. \cite{escande} had recourse to the alternative stress tensor $\ten{P}$ and the artificial bulk force term $\frac{\mu-\lambda}{\rho} \partial_\alpha \rho$ in their LBM algorithms:
the Cauchy stress $\ten{\sigma}$ can not be made to fit the moment chain in (\ref{eq:chainmod}) as elegantly as $\ten{P}$. To model $\ten{\sigma}$ directly, an additional term in (\ref{eq:fieq}) or an additional source term would be required.

Via the third expression of the moment chain
\begin{equation*}
    -\partial_t P_{\alpha \beta} = \partial_\gamma \Big ( \frac{\mu}{\rho} ( j_\alpha \delta_{\beta \gamma} + j_\beta \delta_{\alpha \gamma} + j_\gamma \delta_{\beta \alpha} ) \Big )\,,
\end{equation*}
the stress tensors are related by 
\begin{align}
    -\partial_t P_{\alpha \beta} &
    = \partial_t \Big ( \mu ( \partial_\beta u_\alpha + \partial_\alpha u_\beta + \partial_\gamma u_\gamma \delta_{\beta \alpha} ) \Big )  \nonumber \\ &
    = \partial_t \Big ( \lambda \partial_\gamma u_\gamma \delta_{\alpha \beta}
    + \mu ( \partial_\beta u_\alpha + \partial_\alpha u_\beta) \nonumber- (\lambda-\mu) \partial_\gamma u_\gamma \delta_{\alpha \beta}
    \Big ) \nonumber\\ &
    = \partial_t \Big ( \sigma_{\alpha \beta} - (\lambda-\mu) \partial_\gamma u_\gamma \delta_{\alpha \beta}
    \Big )\,,
\end{align}
and, assuming an adequate initialisation\footnote{The initialisation must fulfil $P_{\alpha \beta}(x,0) = - \sigma_{\alpha \beta}(x,0) + ( \lambda - \mu ) \partial_\gamma u_\gamma(x,0) \delta_{\alpha \beta}$.}
\begin{equation}\label{eq:poissonstress}
    P_{\alpha \beta} = - \sigma_{\alpha \beta} + ( \lambda - \mu ) \partial_\gamma u_\gamma \delta_{\alpha \beta}\,,
\end{equation}
i.e. the stress tensor $\ten{P}$ is equivalent to a negated Cauchy stress tensor $\ten{\sigma}$, with an additional displacement divergence term with coefficient $\lambda-\mu$. The latter disappears for Poisson solids with $\mu=\lambda$ (Poisson's ratio $\nu=\nicefrac{1}{4}$), which is why it is termed the \textit{Poisson stress tensor} in the following.

Thus, the Lam\'e-Navier equation - which is commonly used to model the dynamic behaviour of isotropic solids subject to small displacements - can be rephrased as a moment chain with respect to the current density $\rho$, the linear momentum density $\vec{j}$, and the Poisson stress $\ten{P}$, under the assumptions mentioned above. The LBM acts as a solver for moment chains like (\ref{eq:chainmod}), and may therefore be used to tackle isotropic, small-displacement elastodynamics.

\subsection{Rephrasing the boundary conditions}

In accordance with the reformulation of the Lam\'e-Navier equation as a moment chain in $\rho$, $\vec{j}$, and $\ten{P}$, boundary conditions (BCs) must of course also be rephrased in terms of these variables. 
For Dirichlet boundary conditions, this is comparatively straightforward: with the definition of the linear momentum density $\vec{j}=\rho \partial_t \vec{u}$\footnote{This definition, once more, holds only under smallness assumptions which allow convective terms to be neglected.}, boundary values in the linear momentum density can be obtained from boundary values in the displacement
\begin{equation}\label{eq:bcchaindir}
    j(\vec{x},t) = j^*(\vec{x},t) = \rho(\vec{x},t) \partial_t u^*(\vec{x},t), \quad \vec{x} \in \partial \Omega\,.
\end{equation}

With (\ref{eq:poissonstress}), boundary values in the Poisson stress tensor could be computed from boundary values in the Cauchy stress tensor and from values of the displacement at the boundary, via
\begin{align}
    & P_{\alpha \beta}(\vec{x},t) = P_{\alpha \beta}^*(\vec{x},t) =  - \sigma_{\alpha \beta}^*(\vec{x},t) + ( \lambda - \mu ) \partial_\gamma u_\gamma(\vec{x},t) \delta_{\alpha \beta} \quad \vec{x} \in \partial \Omega\,.\nonumber
\end{align}
However, Neumann boundary conditions enforce boundary values in the Cauchy stress vector $\vec{t}^*$ rather than the Cauchy stress tensor $\ten{\sigma}^*$. Furthermore, the displacement $\vec{u}$ does not appear in the moment chain (\ref{eq:chainmod}), and is thus not directly available in the LBM algorithm.

The latter of these issues may be circumvented by integrating the linear momentum density $\vec{j}=\rho \partial_t \vec{u}$ to obtain the displacement $\vec{u}$. Alternatively, one may approximate the divergence of the displacement $\partial_\alpha u_\alpha$ with the first-order accurate kinematic relation
\begin{equation}\label{eq:approxdiv}
    \partial_\alpha u_\alpha \approx \frac{\rho_0-\rho}{\rho_0}\,,
\end{equation}
as demonstrated in appendix \ref{s:a}.

Meanwhile, Neumann boundary conditions can be transformed into a coordinate system normal to the boundary with a rotation matrix $\ten{T}$
\begin{equation*}
    t_\alpha^{*n}(x,t) = T_{\beta \alpha} t_\beta^*(x,t) \quad x \in \partial \Omega \,. 
\end{equation*}
The entries of the orthonormal transformation tensor $\ten{T}$ are simply given by the normal tangential vectors $\vec{e}^n$ and $\vec{e}^t$ of the boundary, i.e. $T_{1\alpha} = e_\alpha^n$ and $T_{2\alpha}=e_\alpha^t$ for the two-dimensional case \cite[p.28]{holzapfel}.

For the two-dimensional case, (\ref{eq:bcneu}) becomes, in this normal coordinate system
\begin{align}
    \begin{pmatrix}
        \sigma_{nn}^n & \tau_{nt}^n \\
        \tau_{nt}^n & \sigma_{tt}^n
    \end{pmatrix}
    \begin{pmatrix}
        1 \\ 0
    \end{pmatrix}
    =
    \begin{pmatrix}
        t_n^{*n} \\ t_t^{*n}
    \end{pmatrix}\,. \nonumber
\end{align}
As is apparent, the first column of the Cauchy stress tensor $\ten{\sigma}^n$ is fully determined by the Cauchy stress vector to be forced upon the boundary, i.e.
\begin{align}
    \sigma_{nn}^n & = t_n^{*n}\,, \nonumber\\ 
    \tau_{nt}^n & = t_t^{*n}\,, \nonumber
\end{align}
meaning that, with the symmetry of the Cauchy tensor, the Neumann boundary condition determines all entries of $\ten{\sigma}^n$ save for $\sigma_{tt}^n$. This remaining entry remains unaffected by the Neumann boundary condition and may instead be extrapolated from the material behaviour in the vicinity of the relevant boundary.

Thus, we obtain boundary values for the Cauchy stress $\ten{\sigma}^{*n}$ in the normal coordinate system, and an inverse transformation yields boundary values for $\ten{\sigma}^*$ in the original coordinate system,
\begin{equation}\label{eq:trafocauchy}
    \sigma_{\alpha \beta}^*(x,t) = T_{\alpha \gamma} \sigma_{\gamma \zeta}^{*n}(x,t) T_{\beta \zeta}\quad x \in \partial \Omega\,.
\end{equation}
Using (\ref{eq:poissonstress}), we finally obtain the desired boundary values for the Poisson stress tensor $\ten{P}$
\begin{align}
    & P_{\alpha \beta}(\vec{x},t) = P_{\alpha \beta}^*(\vec{x},t) = - \sigma_{\alpha \beta}^*(\vec{x},t) + ( \lambda - \mu ) \partial_\gamma u_\gamma(\vec{x},t) \delta_{\alpha \beta} \quad \vec{x} \in \partial \Omega\,.\label{eq:bcchainneum}
\end{align}
where $\ten{\sigma}^*$ is determined via (\ref{eq:trafocauchy}) and the displacement divergence may be obtained from (\ref{eq:approxdiv}). With (\ref{eq:bcchaindir}) and (\ref{eq:bcchainneum}), Dirichlet and Neumann boundary conditions for elastodynamic problems described by the Lam\'e-Navier equation can now be stated consistently with the moment chain form, completing the problem transformation which allows the LBM to tackle elastodynamic problems.

\section{The solid LBM in a nutshell}\label{s:lbm}

For the sake of compactness and to make this work self-contained, we resort to a very brief outline of the LBM's key features here.
An excellent in-depth discussion can be found in \cite{krueger}.
In keeping with \cite{farag}, we treat the Lattice Boltzmann algorithm as a (rather clever) numerical solver for PDEs described by moment chains, independently from any associations with kinetic theory\footnote{In this view, the distribution functions $\bar{f}_i$ with which the LBM works are merely supplementary variables containing information about the tensors to be modelled (the current mass density $\rho$, linear momentum density $j$, and Poisson stress $P$, in this case).}. 

\begin{figure}[h]
\centering
\includegraphics{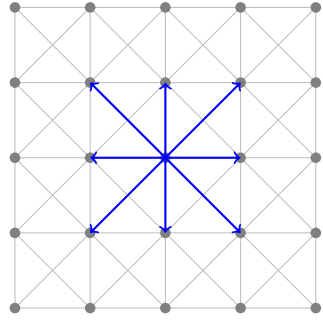}
\caption{(Part of) a D2Q9 lattice for the LBM.}
\label{fig:lattice}
\end{figure}

The algorithm operates on a lattice consisting of regularly spaced lattice sites $\vec{x} \in L \subset \Omega$ (the dark grey points in figure~\ref{fig:lattice}) which are in turn connected by lattice links (the light grey lines in the same diagram) \cite[p.94]{krueger}. Furthermore, several lattice velocity vectors $\vec{c}_i, i \in 0,..,q-1$ (indicated by the blue arrows in figure~\ref{fig:lattice}) cover the distance between any lattice site $\vec{x}$ and its neighbours $\vec{x}+\vec{c}_i \Delta t$ along a lattice link $i$ in one time step $\Delta t$ \cite[p.94]{krueger}. The simplicity of this spatial discretisation makes for easy pre- and post-processing. 

Generally, LBM discretisations are classed by the dimension $d$ and the number of lattice velocities $q$, as DdQq lattices. For ease of visualisation and validation, we only consider D2Q9 lattices - exemplified in figure~\ref{fig:lattice} - here. Note that among the $q=9$ lattice velocities, there is a zero velocity $\vec{c}_0=\vec{0}$.

The LBM encapsulates information about the tensors to be modelled (the current density $\rho$, linear momentum density $\vec{j}$, and Poisson stress $\ten{P}$ in our case) in distribution functions. 
On each site of the lattice, a distribution function vector with one entry $\bar{f}_i$ per lattice velocity $\vec{c}_i$ is introduced \cite[p.63]{krueger}. Here and in the following, we utilise distribution functions resulting from the second-order accurate discretisation by He et al. \cite{he}, which is denoted by the overbar $\bar{\cdot}$.

In each iteration, the distribution functions $\bar{f}_i$ at each lattice site are locally relaxed toward the value of the equilibrium distribution function $f_i^{\text{eq}}$ \cite[p.64]{krueger}, and a contribution due to a source term is added \cite[p.239]{krueger}. The choice of equilibrium distribution function determines the PDE modelled by the LBM algorithm - as shown in \cite{murthy} and \cite{escande}, the expression
\begin{align}
    & f_i^{\text{eq}} = w_i \Big (\rho + \frac{1}{c_s^2} c_{i \alpha} j_\alpha + \frac{1}{2c_s^4} ( P_{\alpha \beta} - \rho c_s^2 \delta_{\alpha \beta} ) ( c_{i \alpha} c_{i \beta} - c_s^2 \delta_{\alpha \beta} ) \Big )\label{eq:fieq}
\end{align}
may be used to recover the Lam\'e-Navier equation. Here, $c_s=\sqrt{\nicefrac{\mu}{\rho}}$ denotes the speed of shear waves, and $w_i$ is a lattice weight associated with lattice link~$i$~\cite{escande}.
The source term
\begin{equation}\label{eq:sourceterm}
    \psi_i = \frac{1}{c_s^2} w_i c_{i\alpha} S_\alpha\,,
\end{equation}
meanwhile, is used to model bulk forces, where 
\begin{equation}
    S_\alpha = F_\alpha + \frac{\mu-\lambda}{\rho} \partial_\alpha \rho
\end{equation}
accounts both for the volumetric loads $\vec{F}$ and the part of the material law not contained in the Poisson stress tensor \cite{escande} - recall the moment chain (\ref{eq:chainmod}).

In the solid LBM algorithm by Murthy et al. \cite{murthy} and Escande et al. \cite{escande}, a BGK collision operator with relaxation time $\bar{\tau}$ is utilised. Together with the contribution due to the source term, this results in the following expression for the post-collision (i.e., post-relaxation) distribution functions \cite[p.239]{krueger}
\begin{equation}\label{eq:coll}
    \bar{f}_i^{\text{col}} = \bar{f}_i - \frac{\Delta t}{\bar{\tau}}( \bar{f}_i - f_i^{\text{eq}} ) + \Delta t \Big (1 - \frac{\Delta t}{2 \bar{\tau}} \Big ) \psi_i\,,
\end{equation}
where the value of $\bar{\tau}$ is crucial to the stability of the method (and the accuracy of boundary condition modelling), with Escande et al. \cite{escande} using $\bar{\tau}=0.55\Delta t$. The locality of this collision step makes parallelisation relatively easy \cite[p.579]{krueger}, contributing to the efficiency which makes the LBM attractive as a PDE solver.

The post-collision distribution functions are subsequently propagated to neighbouring lattice sites along the associated lattice velocities $\vec{c}_i$ in the streaming step \cite[p.66]{krueger}
\begin{equation}\label{eq:stream}
    \bar{f}_i(\vec{x}+\vec{c}_i \Delta t,t+\Delta t) = \bar{f}_i^{\text{col}}\,.
\end{equation}  
Together, the collision and streaming steps define the Lattice Boltzmann equation
\begin{align}
    &\bar{f}_i(\vec{x}+\vec{c}_i \Delta t,t+\Delta t) =  \bar{f}_i - \frac{\Delta t}{\bar{\tau}}( \bar{f}_i - f_i^{\text{eq}} ) + \Delta t \Big (1 - \frac{\Delta t}{2 \bar{\tau}} \Big ) \psi_i\,,\label{eq:LBE}
\end{align}
which describes the evolution of the distribution functions on the lattice \cite[p.239]{krueger}. From the distribution functions, the desired tensors may finally be post-processed as moments
\begin{align}
    \rho&=\sum_i \bar{f}_i\,, \nonumber\\
    j_\alpha & = \sum_i \bar{f}_i c_{i \alpha} + \frac{1}{2} \Delta t S_\alpha\,, \nonumber\\
    P_{\alpha \beta} & = \sum_i \bar{f}_i c_{i \alpha} c_{i \beta}\,.\label{eq:moments}
\end{align}
Consult \cite{farag} for an enlightening discussion of why this works.

\section{Local boundary rules for the solid LBM}\label{s:bound}

\begin{figure}[h]
\centering
\includegraphics{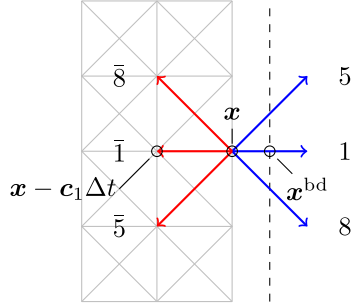}
\caption{Lattice-conforming boundary (dotted line) with lattice velocities leading out of the domain (blue) and lattice velocities for which distribution functions are missing (red) for a boundary lattice site at $\vec{x}$.}
\label{fig:bound}
\end{figure}

While the LBE accounts for the evolution of the distribution functions in the interior of the lattice, the collision and streaming steps leave some distribution functions in the vicinity of the boundary $\partial \Omega \subset \Omega$ undefined. As illustrated in figure~\ref{fig:bound}, these need to be `streamed' from across the boundary to each boundary lattice site $\vec{x} \in \partial L$, i.e., each lattice site with lattice links crossing the boundary \cite[p.165]{krueger}. For convenience, let the set of boundary lattice sites $\partial L \subset L$ be divided into $\partial L_u$ and $\partial L_t$ depending on whether the nearby boundary is subject to Dirichlet or Neumann boundary values, respectively\footnote{Distinctions between boundaries can also be made on a link-by-link basis (as is done in the numerical experiments in section \ref{s:valid}), but this text uses the simpler site-by-site approach to simplify explanations.}. 

Specifying the missing distribution functions in accordance with the boundary conditions of the underlying PDE is a non-trivial task: while boundary conditions are formulated in terms of the tensors appearing in the moment chain ($\rho$, $\vec{j}$, $\ten{P}$), a boundary rule specification with respect to $\bar{f}_i$ is sought \cite[p.155]{krueger}. Furthermore, the number of unknown distribution functions $\bar{f}_i$ at some $\vec{x} \in \partial L$ is frequently higher than the number of boundary condition equations available to specify them, leaving boundary rules for the LBM under-determined \cite[p.155]{krueger}. Additionally, the non-trivial reconciliation of regular lattices with arbitrary bounding geometries complicates work with BCs \cite{jahanshaloo}.

\subsection{Lattice-conforming boundaries}

As a first step, Escande et al. \cite{escande} appropriated the popular bounce-back \cite[p.175]{krueger} and anti-bounce-back \cite[p.200]{krueger} rules from the fluid case, to model lattice-conforming fixed and free boundaries respectively. 
In the scope of this work, these rules were extended, to approximate the behaviour in the vicinity of boundaries with arbitrary Dirichlet and Neumann boundary conditions.

The modified bounce-back rule
\begin{equation}\label{eq:bbcon}
    \bar{f}_{\Bar{i}}(\vec{x},t+\Delta t) = \bar{f}_i^{\text{col}}(\vec{x},t) - \frac{2}{c_s^2} w_i c_{i \alpha} j_\alpha^{*}\,,
\end{equation}
first reverses the direction of motion for the post-collision distribution functions $\bar{f}_i^{\text{col}}(\vec{x},t)$ which would be streamed out of the bulk: the missing distribution functions $\bar{f}_{\Bar{i}}(\vec{x},t+\Delta t)$ for directions $\bar{i}$ are set to the values of the post-collision distribution functions $\bar{f}_i^{\text{col}}(\vec{x},t)$ associated with the opposite direction $i$. Then, a term depending on the linear momentum density boundary values $\vec{j}^*$ is subtracted, and the distribution functions are sent back into the interior of the lattice, in direction $\bar{i}$.
The distribution function $\bar{f}_{\bar{5}}(\vec{x},t+\Delta t)= \bar{f}_7(\vec{x},t+\Delta t)$ in figure~\ref{fig:bound}, for example, is set to the value of the post-collision distribution function $\bar{f}_5^{\text{col}}(\vec{x},t)$, minus the boundary value term $\frac{2}{c_s^2}w_5c_{5\alpha} j_\alpha^*$. 

The required boundary values in the linear momentum density $\vec{j}^*$ can be obtained from Dirichlet boundary values in the displacement $\vec{u}^*$ via (\ref{eq:bcchaindir}). In \cite[p.208-210]{krueger} a derivation of the bounce-back boundary rule in the Chapman-Enskog framework is outlined.
Notably, the change in the distribution functions $\bar{f}_{\bar{i}}$ from one iteration to the next is assumed to be negligible in this derivation, which may give rise to unexpected behaviours when highly transient phenomena in the vicinity of the boundary are considered.

The modified anti-bounce-back boundary rule - used here to model Neumann boundaries - operates similarly. The sign of the post-collision distribution function $\bar{f}_i^{\text{col}}(\vec{x},t)$ at $\vec{x} \in \partial L_t$ is first inverted and a contribution depending on the current density $\rho^{\text{bd}}$ and the Poisson stress $\ten{P}^{*}$ at the boundary is added, before the distribution functions are sent back into the direction from whence they came
\begin{align}
    & \bar{f}_{\Bar{i}}(\vec{x},t+\Delta t) = -\bar{f}_i^{\text{col}}(\vec{x},t) + 2 w_i \Big ( \rho^{\text{bd}} + \frac{1}{2 c_s^4}( P_{\alpha \beta}^{*} - \rho^{\text{bd}} c_s^2 \delta_{\alpha \beta} ) ( c_{i \alpha}c_{i \beta} - c_s^2 \delta_{\alpha \beta} ) \Big )\,.\label{eq:abbcon}
\end{align}
While in \cite{escande}, the current density and Poisson stress are extrapolated normal to the boundary, we use (\ref{eq:bcchainneum}) to determine boundary values in the Poisson stress tensor $\ten{P}^*$ from Neumann boundary conditions on the Cauchy traction vector $\vec{t}^*$. The current density at the boundary $\rho^{\text{bd}}$, meanwhile, is extrapolated along the current lattice link $i$ to $\vec{x}^{\text{bd}}=\vec{x}+\frac{1}{2}\vec{c}_i\Delta t$, 
\begin{equation*}
    \rho^{\text{bd}} = \frac{1}{2} \Big ( 3 \rho(\vec{x},t) - \rho(\vec{x}-c_i \Delta t,t) \Big ) \approx \rho(\vec{x}^{\text{bd}},t) \,.
\end{equation*}
Figure~\ref{fig:bound} indicates, using $i=1$ as an example, how the location of the boundary lattice site $\vec{x}$, neighbouring lattice site $\vec{x}-\vec{c}_i \Delta t$, and boundary site $\vec{x}^{\text{bd}}=\vec{x}+\frac{1}{2}\vec{c}_i \Delta t$ are defined in the lattice-conforming case.

The (admittedly hardly intuitive) link between the anti-bounce-back rule and Neumann boundary conditions may also be derived in the Chapman-Enskog framework. The assumptions and caveats entering this derivation are analogous to those relevant for the bounce-back case.

\subsection{Non-lattice-conforming boundaries}

\begin{figure*}[h]
\centering
\includegraphics{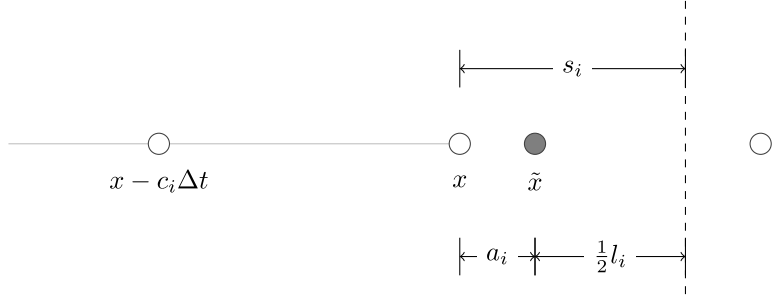}
\caption{Non-lattice conforming boundary for one lattice link with length $l_i$. Adapted from \cite{bouzidi}, figure by the authors.}
\label{fig:boundnoncon}
\end{figure*}

Additionally, an extension to arbitrary geometries is accommodated using the method suggested by Bouzidi et al. \cite{bouzidi} for the fluid case. For a boundary lying at a distance $s_i$ from the boundary lattice site along a lattice link as indicated in figure~\ref{fig:boundnoncon}, this results in 
\begin{align}
    & \bar{f}_{\Bar{i}}(\vec{x},t+\Delta t) = \bar{f}_i^{\text{col}}(\vec{x},t) + \frac{l_i - 2s_i}{l_i + 2s_i} \Big (  \bar{f}_i^{\text{col}}(\vec{x}-\vec{c}_i\Delta t,t) - \bar{f}_{\Bar{i}}^{\text{col}}(\vec{x},t) \Big ) 
    - \frac{2 l_i}{l_i + 2s_i} \frac{2}{c_s^2} w_i c_{i \alpha} j_\alpha^{*}\,,\label{eq:bbnoncon}
\end{align}
for Dirichlet- and 
\begin{align}
    & \bar{f}_{\Bar{i}}(\vec{x},t+\Delta t)  = -\bar{f}_i^{\text{col}}(\vec{x},t) - \frac{l_i - 2s_i}{l_i + 2s_i} \Big ( \bar{f}_i^{\text{col}}(\vec{x}-\vec{c}_i\Delta t,t) + \bar{f}_{\Bar{i}}^{\text{col}}(\vec{x},t) \Big ) \nonumber \\ &
    \quad + \frac{2 l_i}{l_i + 2s_i} 2 w_i \Big ( \rho^{\text{bd}}  + \frac{1}{2c_s^4} ( P_{\alpha \beta}^{*} - c_s^2 \rho^{\text{bd}} \delta_{\alpha \beta} )( c_{i \alpha} c_{i \beta} -c_s^2 \delta_{\alpha \beta} ) \Big )\,,\label{eq:abbnoncon}
\end{align}
for Neumann boundaries, where $l_i$ is the length of lattice link $i$, while $\rho^{bd}$ denotes the linearly interpolated current density
\begin{equation}
    \rho^{\text{bd}} = \frac{l_i + s_i}{l_i} \rho(\vec{x},t) - \frac{s_i}{l_i} \rho ( \vec{x}-\vec{c}_i \Delta t,t)\,.\nonumber
\end{equation}
The caveats mentioned for the lattice-conforming case remain relevant, with the additional limitations of the first-order accurate interpolation. Therefore, the use of finely spaced lattices is imperative when modelling material domains with complicated shapes using (\ref{eq:bbnoncon}) and (\ref{eq:abbnoncon}). Improvements may be attained via quadratic interpolation schemes, but these would come at the cost of reduced efficiency.

\section{Numerical and analytical validation}\label{s:valid}

As demonstrated by Farag et al. \cite{farag}, the Lattice Boltzmann Method should act as a second-order Crank-Nicolson scheme for the moment chain in (\ref{eq:chainmod}), which suggests that the LBM can be used as a numerical solver for problems described by the Lam\'e-Navier equation, Dirichlet- and Neumann boundary conditions, and bulk forces.

This section considers two benchmark examples to verify whether the LBM algorithm and the boundary rules presented above can be used to approach elastodynamic problems in practice.
In the latter of these examples, we consider discontinuous behaviour in time and space to explore the limits of the resulting simulation method.

As our aim is to explore the validity of the LBM as a PDE solver for boundary value problems on the Lam\'e-Navier equation, we do not consider the behaviour of solids observed in experiments, but instead that of ideal (linear) elastic continua as described by analytical and numerical solutions of the underlying PDE. However, the parsimonious anticipation of real-world phenomena is the fundamental objective of solid mechanics, and this section in some sense performs a verification to this end.

\subsection{Tension loading of a plate with a circular hole}

\begin{figure}[h]
\centering
\includegraphics{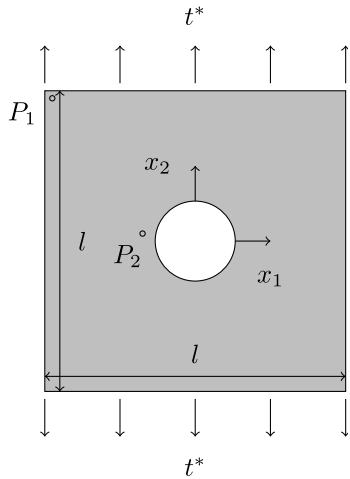}
\caption{Square plate with circular hole subjected to an in-plane tension load.}
\label{fig:tension}
\end{figure}

Firstly, we consider a rectangular plate (side length $l$) with a circular hole (radius $r=0.133l)$, as illustrated in figure~\ref{fig:tension}\footnote{We non-dimensionalised this problem with $u_{\text{ref}}=10^{-3}l$ (arbitrary units).}. This plate is subjected to a plane strain state, the upper and lower boundaries being loaded with a traction of $t^*$ in their respective normal directions. 

This traction is increased linearly from $t^*=0.0\mu \nicefrac{u_{\text{ref}}}{l}$ at time $t=0.0\nicefrac{l}{c_s}$ to $t^*=5.0\mu \nicefrac{u_{\text{ref}}}{l}$ at $t=1.0\nicefrac{l}{c_s}$, and subsequently held at this level until the end of the simulation at $t=2.0\nicefrac{l}{c_s}$. The left and right sides of the plate as well as the circular hole are stress-free, i.e. subject to homogeneous Neumann boundary conditions. 

To describe the simulated material, we use both parameters corresponding to a Poisson solid ($\nicefrac{c_d^2}{c_s^2}=3.0$, $\nu=0.25$) and a non-Poisson solid ($\nicefrac{c_d^2}{c_s^2}=2.8$, $\nu=0.\bar{2}$). The former is often encountered in seismological wave modelling \cite{obrien} and the latter is typical for certain concretes and glasses \cite{mott}.

Lattice Boltzmann simulations with both sets of material parameters and the loads and geometry discussed above are run using a lattice spacing of $\Delta x=0.0125l$. The time step is determined to satisfy the lattice isotropy conditions via \cite[p.64]{krueger}
\begin{equation*}
    \frac{1}{\sqrt{3}} \frac{\Delta x}{\Delta t} = c_s\,,
\end{equation*}
yielding $\Delta t \approx 7.217\times 10^{-3}\nicefrac{l}{c_s}$. A relaxation time of $\bar{\tau}=0.55\Delta t$ is chosen, as in \cite{escande}.

\begin{figure}[h]
\centering
\includegraphics[width=0.45\textwidth]{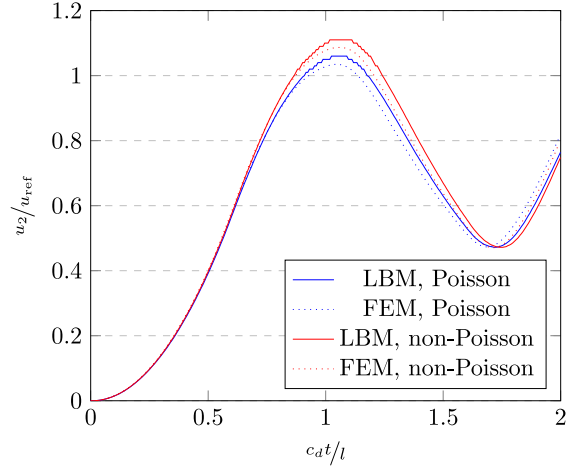}
\caption{$x_2$-components of dimensionless displacements at $P_1$ post-processed from LBM (blue) and FE (red) simulations of a Poisson- and non-Poisson solid under tension loading.}
\label{fig:tensiony}
\end{figure}

After the simulation run, the $x_2$-component $u_2$ of the displacement at the point $P_1=(-0.5l+\nicefrac{\Delta x}{2},0.5l-\nicefrac{\Delta x}{2})$ in the top-left corner and the x-component $u_x$ of the displacement at the point $P_2=(-0.15l+\nicefrac{\Delta x}{2},\nicefrac{\Delta x}{2})$ near the hole are post-processed from the linear momentum density $j$. The results are compared with those produced by a transient Finite Element (FE) simulation with an equivalent geometry, material parameters, loads, and discretisation\footnote{The FE analyses were run using the FEAP FE program (see e.g. \cite{feap}), using standard Newmark time integration and quad elements with bilinear shape functions.}. 

Figure~\ref{fig:tensiony} plots the vertical displacement ($u_2$) predicted at $P_1$ by the LBM and the FEM for the Poisson case in blue, and the non-Poisson case in red. The reference results produced by the FEM are displayed as dotted lines, while the continuous curves indicate the LBM's prognosis. 

As is apparent in the figure, the LBM and the FEM are in good agreement throughout the simulation. This indicates that the solid LBM with the modified anti-bounce-back rule captures the transient behaviour of the Lam\'e-Navier equation with (continuous) Neumann boundary values well in this example. This agreement is achieved even though, strictly speaking, the tractions $t^*$ violate the assumption of quasi-static loading made in the derivation of the boundary rules. The error introduced by this assumption seems sufficiently small in this case for the desired solution to be obtained regardless. The boundary rules for both the lattice-conforming and non-lattice-conforming boundaries appear to work as desired.

\begin{figure}[h]
\centering
\includegraphics[width=0.45\textwidth]{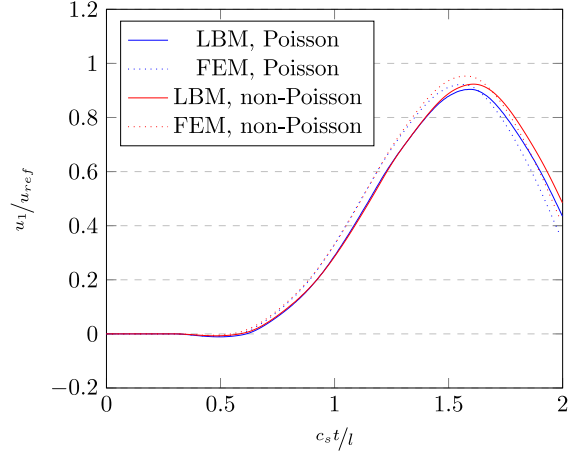}
\caption{$x_1$-components of dimensionless displacements at $P_2$ postprocessed from LBM (blue) and FE (red) simulations of a Poisson- and a non-Poisson solid under tension loading.}
\label{fig:tensionx}
\end{figure}

Similarly, figure~\ref{fig:tensionx} visualises the change of horizontal displacement ($u_2$) at $P_2$ computed via the LBM and the FEM. As for the Poisson solid above, the displacements are in good agreement throughout the simulation. Interestingly, the LBM seems to predict a slightly later peak in the displacement $u_2$. This may be partially due to the well-known offset in the effective location of a boundary simulated using bounce-back boundary rules \cite[p.210]{krueger}. The first-order interpolation used for the non-lattice-conforming boundaries may further exacerbate this effect. However, the error produced this way is not significant, and the LBM captures the behaviour predicted by the FEM both qualitatively and quantitatively.

\begin{figure*}[h]
\centering
\includegraphics[trim=0 30 0 30,clip,width=\textwidth]{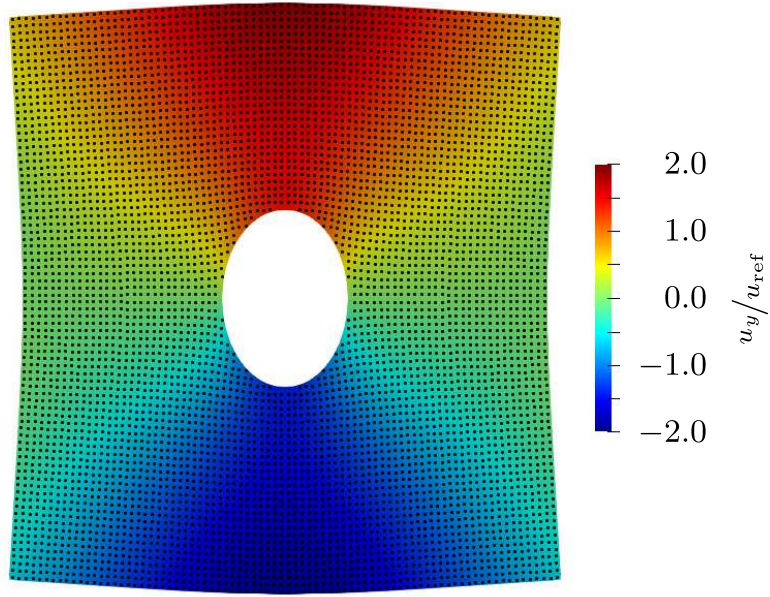}
\caption{Deformed configurations predicted at $t=1.5\nicefrac{l}{c_s}$ by FEM and LBM simulations. A contour plot for the vertical displacement $u_2$ is superimposed on the warped FEM domain, while the lattice sites of the LBM simulation are indicated by black dots. The deformation is scaled by a factor of $20$.}
\label{fig:defo}
\end{figure*}

Finally, the deformation predicted by the LBM and FEM simulations at $t=1.5\nicefrac{l}{c_s}$ is visualised in figure~\ref{fig:defo}, for the non-Poisson solid with $\nicefrac{c_d^2}{c_s^2}=2.8$. A contour plot for the vertical displacements $u_2$ is superimposed on the warped surface indicating the current configuration as determined by the FEM. The position of lattice sites in the current configuration is indicated by black dots. The agreement in the computed displacements throughout the simulated domain is apparent.

The agreement between the two simulation methods in this example is encouraging. With the modified anti-bounce-back rule, the LBM seems to capture elastodynamic behaviour roughly as well as the FEM. Though instabilities are a common issue with the BGK-LBM \cite{escande}, the transient LBM simulations here remain stable for a sufficiently long duration to capture the transient overshoots in the displacement of the dynamically loaded solid without notable issues. Furthermore, the additional linear momentum density source - see (\ref{eq:chainmod}) - seems to model the non-Poisson part of the material law successfully in this application.

\subsection{Sudden loading of a stationary crack}

\begin{figure}[h]
\centering
\includegraphics{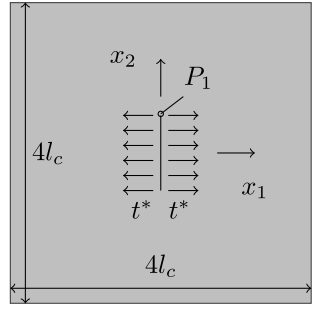}
\caption{Square plate with a crack (length $l_c$) subjected to tensile crack face traction. Side length not to scale.}
\label{fig:crack}
\end{figure}

In the second numerical example, the LBM is used to treat a problem of dynamic fracture mechanics: the square cross-section\footnote{The finite material domain modelled here is sufficiently large for our purposes: no P waves reflected by the exterior boundaries can return to and influence the crack within the simulated time.} in figure \ref{fig:crack} with side length $l=4.0l_c$ and free boundaries features a stationary crack of length $l_c$\footnote{We non-dimensionalised this problem with $l_{\text{ref}}=l_c=1.0$ (arbitrary units). The odd-looking non-dimensionalised quantities follow from our choice of material parameters, namely $\mu=1.3$.}. The crack is oriented in the y-direction and centred at the origin $(0\nicefrac{l}{l_{c}},0\nicefrac{l}{l_{c}})$. Both crack faces are suddenly loaded with a traction $t^*=0.009615\mu\nicefrac{u_{\text{ref}}}{l}$ in their negative normal directions at time $t=0.0\nicefrac{l}{c_d}$, and the traction remains constant until the end of the simulation at $t=2.0\nicefrac{l}{c_d}$. The LBM simulates the consequences of this sudden so-called `mode I' loading of the crack for a non-Poisson solid with $\nicefrac{c_s^2}{c_d^2}=0.3611$, $\nu=0.2174$. 
A spatial discretisation with $\Delta x=0.01l_c$ is chosen, leading to a time-step of $\Delta t \approx2.669\times 10^{-3}\nicefrac{l}{c_d}$. We further use a relaxation time of $\bar{\tau}=0.55\Delta t$. 

\begin{figure}[h]
\centering
\includegraphics{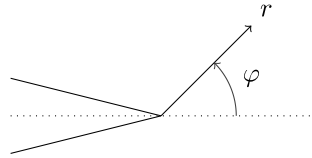}
\caption{Cylindrical coordinate system at the crack tip and definition of the displacement. Adapted from \cite[p.71]{becker}.}
\label{fig:cracktip}
\end{figure}

Following the simulation, the $x_1$-component of the stress $\sigma_{11}$ at a point $P_1=(0.0l_c,0.52l_c)$ just in front of one of the crack tips is evaluated and used to calculate the mode I stress intensity factor $K_I$ for each time step via \cite[74]{gross}
\begin{equation}
    K_I = \lim_{r\rightarrow0} \sigma_{11}(r,\varphi) \sqrt{2 \pi r}\,.
\end{equation}
Here, $r$ is the distance from the crack tip to $P_1$ and $\varphi=0$ the angle with which $P_1$ is offset from the crack plane (see figure~\ref{fig:cracktip}).

The results are compared to the analytical solution from Freund \cite[p.117]{freund}, which yields
\begin{equation}
    K_I^0 = 2t^* \frac{\sqrt{1-2\nu}}{1-\nu} \sqrt{\frac{c_d t}{\pi}}\,,
\end{equation}
for the mode-I stress intensity factor from $t=0.0\frac{l_c}{c_d}$ to $t=1.0\frac{l_c}{c_d}$, i.e. until the first dilatational waves scattered by one crack tip arrive at the other \cite[p.119]{freund}.
The stress intensity factor from $t=\frac{l_c}{c_d}$ to $t=2\frac{l_c}{c_d}$ (i.e., until the dilatational waves return to the other crack tip once more) is given by the nested integral \cite[p.122]{freund}
\begin{align}
    & K_I^1 = K_I^0 +\frac{2 t^* F_+(0)}{\pi} \sqrt{\frac{2l_c}{\pi}} \int_a^{\frac{t}{l_c}} \operatorname{Im} \Big ( \sqrt{\frac{t}{l_c}-\eta} \sqrt{\frac{a-\eta}{a+\eta}} \frac{c_r+\eta}{c_r-\eta} \frac{1}{\eta} \exp \Big ( \cdot \cdot \nonumber \\
    & \quad \frac{1}{\pi} \int_a^b \arctan \Big ( 4 \xi^2 \frac{\sqrt{(\xi^2-a^2)(b^2-\xi^2)}}{(b^2-2 \xi^2)^2} \Big ) \frac{2 \eta}{\xi^2-\eta^2} d \xi \Big ) d \eta \Big )\,,\label{eq:ki1}
\end{align}
with $a=\frac{1}{c_d}$ and $b=\frac{1}{c_s}$.
$c_r=\nicefrac{1}{c}$ is the inverse of the Rayleigh (surface) wave speed, estimated as $c_r\approx\nicefrac{b(1+\nu)}{0.862+1.14\nu}$ in \cite[p.83]{freund}. To increase the robustness of the solution over a wider range of parameters, we instead compute the surface wave speed via the roots of the equation \cite{rahman}
\begin{equation*}
    0 = \Big ( 2 - \Big ( \frac{c}{c_s} \Big )^2 \Big )^2 - 4 \sqrt{ \Big ( 1 - \Big ( \frac{c}{c_d} \Big )^2 \Big ) \Big ( 1 - \Big ( \frac{c}{c_s} \Big )^2 \Big )}\,.
\end{equation*}
The meaning of the function $F_+(\xi)$ is explained in \cite[p.90]{freund}. Here, it suffices to know that it assumes the value of $F_+(0)\approx 0.8774$ in (\ref{eq:ki1}). 

Crucially, the inner and outer integrands feature poles at $\xi=\eta$ and $\eta=c_r$, respectively. Since the integrands are sufficiently well-behaved and change sign as the variable of integration crosses the pole, Cauchy principal values exist \cite[p.122]{freund}. Here, we evaluate (\ref{eq:ki1}) using a trapezoidal rule with an adaptive integration step near the poles, and we integrate around the inner singularity $\xi=\eta$ along a small semi-circular contour in the upper complex half-plane. A tolerance of $2\times10^{-5}$ at the singularities seems sufficient for convergence\footnote{A lower tolerance does not visibly change the computed analytical solution.}. The MATLAB script for the evaluation of the Cauchy principal value integrals is available upon request.

\begin{figure}[h]
\centering
\includegraphics[width=0.45\textwidth]{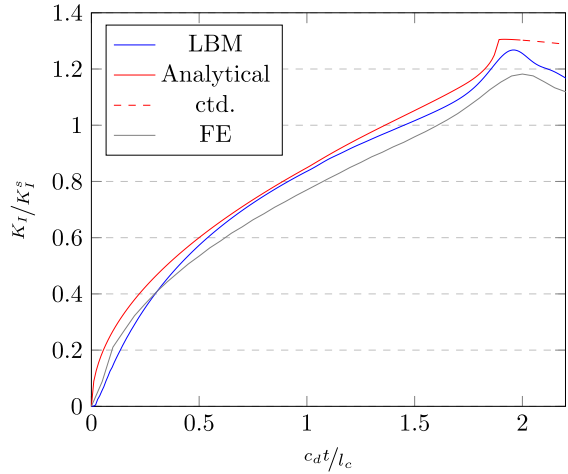}
\caption{Mode-I stress intensity factor postprocessed from the LBM simulation of a non-Poisson solid with a shock-loaded crack at $P_1=(0.005,0.52)$, compared with an analytical solution calculated using the Wiener-Hopf method. Stress intensity factor normalised with the quasistatic solution $K_I^s = t^* \sqrt{\nicefrac{\pi l_c}{2}}$.}
\label{fig:pcrack}
\end{figure}

The numerically integrated analytical solution for $K_I$ is plotted against the results postprocessed from the LBM simulation run at $P_1$ in figure~\ref{fig:pcrack}. The analytical solution is not exact after $t=2\frac{l_c}{c_d}$, and the plot is continued with a dashed line. The exact solution, however, is expected to deviate only slightly from this dashed line \cite[p.123]{freund}. The LBM produces results that generally agree well with the analytical solution and a reference FE calculation: the stress intensity factor rises steeply initially, with the rate of change decreasing gradually. After $t=\frac{l_c}{c_d}$ the solutions both behave almost linearly in time as stress waves from the other end of the crack arrive, with a notable uptick just before $t=2\frac{l_c}{c_d}$ due to the arrival of Rayleigh surface waves \cite{chirino}. $K_I$ then levels off.
The LBM (with the modified boundary rules) and the FE produce a smoother behaviour than the exact solution. For the LBM, this is to be expected because the bounce-back boundary rules were derived under the assumption of small temporal gradients for the continuum mechanical fields. However, the solution produced by the LBM captures key aspects of the behaviour in the vicinity of the crack tip roughly as well as the FEM, both qualitatively and quantitatively.
Encouragingly, the simulation is stable despite the discontinuous loading and the singular stress field in the vicinity of the crack tip.

\section{Conclusions}

In this work, simple local boundary rules for the LBM algorithm by Murthy et al. \cite{murthy} and Escande et al. \cite{escande} were proposed. These approximate the behaviour of Dirichlet and Neumann boundary conditions, both for the straightforward lattice-conforming case and for the more complicated case of arbitrary geometries.

In section~\ref{s:mech}, the Lam\'e-Navier equation and the associated boundary conditions were first rephrased in moment chain form. The relationship between the Poisson stress tensor $\ten{P}$ appearing in this moment chain and the established Cauchy stress tensor $\ten{\sigma}$ was explored, and a convenient formula for the calculation of one from the other derived. 
Section~\ref{s:lbm} then very briefly outlined the Lattice Boltzmann Method for solids, and section~\ref{s:bound} presented the boundary rules used to model arbitrarily valued Dirichlet and Neumann boundary conditions with arbitrary geometries.
In section~\ref{s:valid}, the LBM for elastic solids and the novel boundary rules were finally validated against one analytical and one numerical benchmark problem.

However, open issues remain. The solid LBM and the modified boundary rules are not unconditionally stable.
Furthermore, the boundary rules are not necessarily valid for all relevant time-dependent boundary values. This may explain the slightly `softer' behaviour exhibited by the LBM for highly transient simulations.
Moreover, the de facto simulated location of the boundary is offset slightly from the intended location, lattice-conforming boundary rules are second-order accurate in space in the best case, and non-lattice-conforming boundary rules only allow for first-order accuracy. 

Beyond this, the algorithm outlined here only facilitates the modelling of materially and geometrically linear solid behaviour. For an extension to the nonlinear regime, nonlinear terms in the material law must be accounted for. Furthermore, a formulation in the reference configuration would be desirable to simplify boundary modelling under large deformations.

On the other hand, the LBM seems promising as an elastodynamic solver: the locality of its nonlinearity and the linearity of its non-locality \cite[p.55]{krueger} make the LBM exceptionally efficient and easily parallelizable. 
In particular, the LBM lends itself to modelling dynamic phenomena which call for fine spatial and temporal resolutions -- an area of application in which computational efficiency becomes crucial.
Furthermore, spatially discretising - `meshing' - a material domain is comparatively easy. If the solid LBM can be turned into a practicable simulation method, much time might be saved in engineering design processes.

The numerical solutions to the dynamic example problems considered here are encouraging: the results produced by the LBM are in good agreement with analytical and numerical benchmark results. 

However, a need for further development is apparent in several key areas. Firstly, the stability of the method must be improved. Schemes with multiple relaxation times \cite[p.407]{krueger} and multi-reflection boundary rules \cite{ginzburg} seem promising first places to start. Furthermore, high-frequency waves could be filtered \cite{escande}, or distribution functions `reset' regularly.
Second-order accurate non-lattice-conforming boundary rules could further improve the performance of the LBM on problems with complicated bounding geometries. Finally, extensions to material nonlinearity and large deformations are desirable. 

If these shortcomings are addressed, however, fascinating areas of application beckon: the LBM could be used to study stresses caused by the propagation, reflection, and superposition of waves under highly dynamic loading. Crashproofing and mining are fields that might benefit from such studies. If an extension to the nonlinear regime can be achieved, simulations of polymer components (such as seals and tires) or tissue (such as muscle) under transient loads further become feasible. Here, also, the LBM might be useful on account of its efficiency and ease of parallelization.

\begin{appendices}

\section{An approximation for the displacement divergence}\label{s:a}

For finite strains, the current density is coupled to the kinematic description of the continuum via the volume ratio $J$, i.e. the determinant of the deformation gradient $F$, \cite[p.74]{holzapfel}
\begin{align}
    J & = \det F = \prod_{\vartheta=1}^{d} F_{(\vartheta\vartheta)}=  \prod_{\vartheta=1}^{d} \frac{\partial x_{(\vartheta)}}{\partial X_{(\vartheta)}} \nonumber \\ & = \prod_{\vartheta=1}^{d} \Big (1+\frac{\partial u_{(\vartheta)}}{\partial X_{(\vartheta)}} \Big) = \frac{dV}{dV_0} = \frac{\rho_0}{\rho}\,,
\end{align}
where $d$ denotes the dimension of the continuum, and $\vartheta$ is the coordinate index in the principal coordinate system (summation over this index is not implied). For $d=3$, expansion yields
\begin{align}
    J = & 1 + \sum_{\vartheta=1}^3 \frac{\partial u_{(\vartheta)}}{\partial X_{(\vartheta)}} \nonumber \\& + \sum_{\iota=1}^3 \sum_{\kappa=1}^{\iota-1} \frac{\partial u_{(\iota)}}{\partial X_{(\iota)}} \frac{\partial u_{(\kappa)}}{\partial X_{(\kappa)}} + \prod_{\vartheta=1}^3 \frac{\partial u_{(\vartheta)}}{\partial X_{(\vartheta)}}\,.\nonumber
\end{align}
Due to the invariance of the trace, the second term may also be written in terms of the original (non-principal) coordinate system, and
\begin{align}
    J = & 1 + \frac{\partial u_{\alpha}}{\partial X_{\alpha}} \nonumber \\& + \sum_{\iota=1}^3 \sum_{\kappa=1}^{\iota-1} \frac{\partial u_{(\iota)}}{\partial X_{(\iota)}} \frac{\partial u_{(\kappa)}}{\partial X_{(\kappa)}} + \prod_{\vartheta=1}^3 \frac{\partial u_{(\vartheta)}}{\partial X_{(\vartheta)}}\,.\label{eq:Jexp}
\end{align}
where we make use of Einstein summation convention again.

Meanwhile, the displacement gradient $\ten{H}=\nicefrac{\partial \vec{u}}{\partial \vec{X}}$ is linked to the derivative of the displacement $\nicefrac{\partial \vec{u}}{\partial \vec{x}}$ in the current configuration $\Omega$ by
\begin{align}
    H_{\alpha \beta} & = \frac{\partial u_\alpha}{\partial X_\beta} = \frac{\partial u_\alpha}{\partial x_\gamma} \frac{\partial x_\gamma}{\partial X_\beta} \nonumber \\&
    = \frac{\partial u_\alpha}{\partial x_\gamma} \Big ( \delta_{\beta \gamma} + \frac{\partial u_\gamma}{\partial X_\beta} \Big )\,.\label{eq:uel}
\end{align}
If the displacement derivatives $\nicefrac{\partial \vec{u}}{\partial \vec{x}}$ and $\nicefrac{\partial \vec{u}}{\partial \vec{X}}$ are sufficiently small, lower-order terms in them dominate terms at a higher order of smallness (see, e.g. \cite[p.3]{landau}). By substituting (\ref{eq:uel}) into (\ref{eq:Jexp}) and considering only terms at the first order of smallness, we obtain
\begin{equation*}
    J = \frac{\rho_0}{\rho} = 1 + \frac{\partial u_{\alpha}}{\partial x_{\alpha}} = 1 + \partial_\alpha u_\alpha\,,
\end{equation*}
and thus
\begin{equation*}
    \partial_\alpha u_\alpha = \frac{\rho_0}{\rho}-1 = - \frac{\rho-\rho_0}{\rho} \approx - \frac{\rho-\rho_0}{\rho_0}\,,
\end{equation*}
where the density fluctuation in the denominator is neglected as $\rho= \nicefrac{\rho_0}{1+\partial_\alpha u_\alpha}$ is dominated by the constant term.

\end{appendices}

\bibliographystyle{unsrtnat}
\bibliography{main}  






\end{document}